# Gold-Gold Bonding: The Key to Stabilizing the 19-Electron Ternary Phases *Ln*AuSb (*Ln* = La-Nd and Sm) as New Dirac Semimetals

Elizabeth M. Seibel[*], Leslie M. Schoop, Weiwei Xie, Quinn D. Gibson, James Webb, Michael K. Fuccillo, Jason W. Krizan, and Robert J. Cava[#]

Department of Chemistry, Princeton University, Princeton, New Jersey 08544, USA

**Abstract**

We report a new family of ternary 111 hexagonal *Ln*AuSb (*Ln* = La-Nd, Sm) compounds that, with a 19 valence electron count, has one extra electron compared to all other known *Ln*AuZ compounds. The "19[th]" electron is accommodated by Au-Au bonding between the layers; this Au-Au interaction drives the phases to crystallize in the YPtAs-type structure rather than the more common LiGaGe-type. This is critical, as the YPtAs structure type has the symmetry-allowed band crossing necessary for the formation of Dirac semimetals. Band structure, density of states, and crystal orbital calculations confirm this picture, which results in a nearly complete band gap between full and empty electronic states and stable compounds; we can thus present a structural stability phase diagram for the *Ln*AuZ (Z = Ge, As, Sn, Sb, Pb, Bi) family of phases. Those calculations also show that LaAuSb has a bulk Dirac cone below the Fermi level. The YPtAs-type *Ln*AuSb family reported here is an example of the uniqueness of gold chemistry applied to a rigidly closed shell system in an unconventional way.

**Introduction**

Ternary 1:1:1 *LnYZ* (*Ln* = lanthanide; *Y* = transition metal; *Z* = main group) phases are of interest because they can exhibit interesting magnetic and electronic properties due to the presence of itinerant transition metal electrons and localized or hybridized rare-earth *f* electrons.[1] These phases crystallize mainly in two crystal systems: one set with the well-known face centered cubic Half-Heusler structure type, and a second set in several hexagonal structure types. The most common hexagonal structure is known as the LiGaGe-type, which involves graphene-like single layers of Ga and Ge ordered in a honeycomb net separated by layers of Li.[1,2] Complex structural variants among the hexagonal phases are also known.[2,3,4] Although formal electron counting can be appreciated for compounds like $Na^{+1}Cl^{-1}$ and $Ga^{+3}As^{-3}$, it is not so obvious in ternary compounds like this. Nonetheless there are strong electron counting rules for chemical stability in the 1:1:1 family addressed here: 18 electrons are required to fill the bonding states, leaving the antibonding states empty and leading to a chemically stable, semiconducting compound with a band gap between the filled and empty states.[1] Therefore, the 1:1:1 compounds of this type are nearly always 18-electron systems (counted as the sum of the rare-earth *s* and *d* electrons, transition metal *s* and *d* electrons, and main group *s* and *p* electrons).[2-3]

The most common hexagonal structure, the LiGaGe-type, requires 18 electrons to fill the bonding states up to the gap at the Fermi level ($E_F$).[1] Typically the introduction of a "19$^{th}$" electron makes the structures chemically unstable and they do not form; here we report here the unusual, 19-electron *Ln*AuSb phases (*Ln* = La-Nd and Sm). These compounds crystallize in a 4-layer structure of the YPtAs[5]-type, which has the symmetry-allowed band crossing necessary for Dirac semimetals. We show that the 19$^{th}$ electron is localized in a molecular-like, interlayer Au-Au dimer bond, resulting in a nearly complete band gap between filled bonding states and empty antibonding states, and thus chemical stability. The compounds should therefore be considered as $Ln^{3+}{}_2(Au-Au)^0Sb^{3-}{}_2$ phases. Their electronic relationship to other *Ln*AuZ phases is described.

**Experimental**

    **Syntheses**

99.9% purity rare earth and > 99.99% purity other metals were used as starting materials. The rare earth elements were arc-melted before use and stored in an inert atmosphere. For $Ln$AuSb compounds, the rare earth, $AuSb_2$ (made by melting at 850 °C), and Au were used as starting materials to minimize Sb vaporization on arcmelting. These starting materials were arcmelted in argon in a 1:1:1 Ln:Au:Sb stoichiometry. Samples were then annealed in evacuated silica tubes for 48 h at 850 °C. Longer annealing times or hotter annealing temperatures caused decomposition into $Ln_3Au_3Sb_4$, and the use of excess Sb during arc melting led to the formation of impurity phases. We also synthesized LaAuSn for comparison purposes (as described below) by arcmelting the elements in an argon atmosphere in a 1:1:1 ratio and then annealing at 850 °C for 1 week. The arcmelted buttons were stable in air for days, but sample grinding was carried out in an inert atmosphere to prevent rapid oxidation.

**X-ray Powder Diffraction**

Samples were initially checked for phase purity by powder X-ray diffraction (pXRD) using a Rigaku MiniFlex II with CuKα radiation and a diffracted beam monochrometer. Synchrotron powder X-ray diffraction data for structure refinement was collected at beam line 11-BM at the Advanced Photon source at Argonne National Laboratory for LaAuSb, CeAuSb, and NdAuSb at 298 K. The resulting powder diffraction patterns were refined using the FullProf suite. Small single crystals of LaAuSb and LaAuSn were selected from the arcmelted buttons for single crystal study. Single crystals were mounted on the tips of glass fibers and room temperature intensity data were collected on a Bruker Apex II diffractometer with Mo radiation $K\alpha_1$ ($\lambda$=0.71073 Å). Data were collected over a full sphere of reciprocal space with 0.5° scans in ω with an exposure time of 20s per frame. The 2θ range extended from 4° to 60°. The SMART software was used for data acquisition. Intensities were extracted and corrected for Lorentz and polarization effects with the SAINT program. Empirical absorption corrections were accomplished with SADABS which is based on modeling a transmission surface by spherical harmonics employing equivalent reflections with I > 2σ(I).[6,7] The crystal structure of LaAuSn was solved using direct methods and refined by full-matrix least-squares on $F^2$ using the SHELLX package.[8] All crystal structure drawings were produced using the program *VESTA*.[9]

**Structure Determinations**

Rietveld refinements of synchrotron X-ray powder diffraction data were carried out for LaAuSb, CeAuSb, and NdAuSb. $Ln_3Au_2Sb_3$, $Ln_{14}Au_{51}$, and $Ln_3Au_3Sb_4$ impurity phases were observed. To the best of our knowledge, of the $Ln_3Au_2Sb_3$ phases only $Ce_3Au_2Sb_3$ has been reported.[10] Because all samples contained at least one impurity phase, the composition of the $Ln$AuSb phases was fixed in the 1:1:1 stoichiometry in the refinements. The atomic positions for YPtAs were used as starting points for the powder refinements; different models were tested with variations of the Au and Sb positions and extent of honeycomb buckling; all refinements quickly converged. Full structure solutions for PrAuSb and SmAuSb were not attempted due to the presence of significant amounts of the $Ln_3Au_3Sb_4$ impurity phase, but their unit cell parameters were easily determined.

The structure of LaAuSn was previously reported as a disordered version of the LiGaGe type with Au/Sn mixing (the $CaIn_2$-type)[11]; however, we find that the compound crystallizes in the ordered, 2-layer LiGaGe-type, in agreement with other $Ln$AuSn phases[12]. Relevant structure parameters refined from single crystal data for LaAuSn are given in Tables S1 and S2.

**Electronic Calculations**

The electronic structure of LaAuSb was calculated with the aid of *CAESAR*[13], according to semi-empirical extended-Hückel-tight-binding (EHTB) methods. The parameters for Au are $6s$: $\zeta = 1.890$, $H_{ii} = -8.23$ eV; $6p$: $\zeta = 1.835$, $H_{ii} = -4.89$ eV, and $5d$: $\zeta = 3.560$, $H_{ii} = -12.200$ eV. The Au parameters were modified to provide the best fit to the results of first-principles calculations with relativistic effect.[14, 15]. Partial Density of States (DOS) and Crystal Orbital Hamilton Population (COHP) calculations[16] were performed by the self-consistent, tight-binding, linear-muffin-tin-orbital (LMTO) method in the local density (LDA) and atomic sphere (ASA) approximations, within the framework of the DFT method.[17,18,19]

Interstitial spheres were introduced in order to achieve space filling. The ASA radii as well as the positions and radii of these empty spheres were calculated automatically, and the values so obtained were all reasonable. Reciprocal space integrations were carried out using the tetrahedron method. Down-folding techniques were automatically applied to the LMTOs, and scalar relativistic effects were included in the calculations.

Further, ab-initio electronic band structure calculations were performed in the framework of density functional theory (DFT) using the *WIEN2k*[20] code with a full-potential linearized augmented plane-wave and local orbitals [FP-LAPW + lo] basis[21, 22, 23] together with the Perdew-Becke-Ernzerhof (PBE) parameterization[24] of the Generalized Gradient Approximation (GGA) as the exchange-correlation functional. The plane wave cut-off parameter $R_{MT}K_{MAX}$ was set to 8 and the Brillouin zone was sampled by 2000 k-points. Experimental lattice parameters from the Rietveld refinements (for LaAuSb and LaAuSn) and from published data (LaAuPb)[25] were used in the calculations. Spin orbit coupling (SOC) and relativistic effects were included.

**Results and Discussion**

### Structural Description.

Among the *Ln*AuSb (*Ln* = La-Nd, Sm) phases, CeAuSb was previously known to exist but its structure was undetermined. The powder X-ray diffraction data was interpreted as indicating a 2-layer, disordered LiGaGe -type structure (Figure 1), but with impurity phases present.[26] In a more recent study of the Ce-Au-Sb phase diagram, the existence of CeAuSb was confirmed by SEM-EDS, but the observed powder diffraction pattern was found to be inconsistent with the previously reported structure.[10] In our investigation of *Ln*AuSb (*Ln* = La-Nd, Sm) phases, we found that a small single crystal selected from an arcmelted button of LaAuSb rather had twice the *c*-axis previously reported for CeAuSb, as expected for the 4-layer YPtAs structure type (Figure 1). Inspection of the CeAuSb powder diffraction pattern then subsequently showed that peaks previously associated with impurity phases, present in all the *Ln*-phases reported here, are in fact captured by a doubling of the *c*-axis, i.e. by a 4-layer rather than a 2-layer structure. Thus we determined that a four layer structural pattern, rather than a two-layer pattern, describes the crystal structures of the light rare earth compounds in the *Ln*AuSb family. The single crystal data was not of sufficient quality to perform a full structural study, and therefore the structures were determined quantitatively from the synchrotron powder X-ray diffraction data. The powder pattern and fit to the data for LaAuSb is shown in Figure 2 as an example (see SI Figures S1 and S2 for CeAuSb and NdAuSb). Table 1 gives the refined structural parameters for LaAuSb, CeAuSb, and NdAuSb, and Table 2 lists the lattice parameters for all the *Ln*AuSb (*Ln* = La-Nd, Sm) phases observed in this study.

Figure 3 shows the refined structure of LaAuSb, and Figure 4 compares this to the archetype YPtAs. In the YPtAs structure type the only variable positional coordinates are the *z*-axis parameters of the Pt and As sites that form the $Y_3Z_3$ honeycomb layers. These parameters reflect the degree of layer buckling and thus the degree of *Y-Y* interlayer bonding; for the current materials this is an Au-Au bond. As can be seen in the figure, the *Ln*AuSb phases are more buckled than the YPtAs prototype, and in that respect are similar to LiGaGe. However, in the LiGaGe structure type the honeycomb layers are arranged so that the *Y* and *Z* atoms are above each other resulting in *Y-Z* interlayer bonding and *Y-Y* interlayer bonding is not possible. However, in the YPtAs structure type observed for our *Ln*AuSb phases, the neighboring honeycomb layers alternate their stacking in a …*YYZZYYZZ*…arrangement and buckle (Figure 1), which we attribute to the Au-Au bond formation. The resulting coordination polyhedra and selected bond lengths for LaAuSb are shown in Figure 4, and Table S3 summarizes the interatomic bond lengths for LaAuSb, CeAuSb, and NdAuSb. In all compounds the rare earth atoms adopt 12-fold coordination, whereas the Sb atoms form a trigonal prism with *Ln*. Three bent Au-Sb bonds cut through this prism. If the Au-Au dimer is treated as one unit, it adopts 9-fold coordination made of a Sb trigonal prism with planar trigonal Ln bonds. The near neighbor coordination polyhedra are generally the same in shape in LiGaGe structure type compounds, but the atoms found at the vertices are different.

Though the cutoff for bonding is somewhat arbitrary, here we consider interactions less than 3.15 Å to be a covalent bond for Au-Au contacts.[2] In $Ln$AuSb ($Ln$ = La-Nd, Sm) the honeycomb layer buckling allows the gold atoms in neighboring planes to approach each other at a distances of 2.98 Å, 3.05 Å, and 3.12 Å for Nd, Ce, and La, respectively (see Table S3 for selected bond distances). In contrast, the intralayer Au-Sb bond length is nearly constant at 2.77-2.78 Å. Thus the primary influence of the size of the rare earth ion is to modulate the spacing between the honeycomb layers. The Au-Au interlayer bonds of ~ 3 – 3.1 Å are longer than the bonding in metallic gold (2.88 Å), but they are squarely within the range of "aurophilic interactions" (2.8-3.5 Å).[27]

The term "aurophilic interaction" is most strictly used to describe the affinity between two closed-shell gold centers ($Au^{1+}$, $5d^{10}$) driven by relativistic effects and the high electronegativity of gold.[27] In molecular systems, these aurophilic interactions can often be significant enough to drive dimerization (e.g. for $[(Me_2PCH_2PMe_2)_3Au_2]^+$) and crystallization (e.g. for 1,1'-di(isocyano)ferrocene).[27] The term is also often applied to mixed-valence interactions (between $Au^{-1}[6s^25d^{10}]$, $Au^0[6s^15d^{10}]$, and/or $Au^{1+}[5d^{10}]$) in molecules, but for these types of open-shell interactions (like $Au^0$-$Au^{1+}$) the aurophilicity may have a smaller impact on cluster formation and a description in terms of metal-metal bonding may be more appropriate.[27] The tendency for gold to form auride anions in the solid state due to its high electronegativity has also previously discussed from a Zintl perspective for compounds like CsAu and $Cs_3AuO$.[28] Weak gold-gold interactions have been observed in other hexagonal 111 phases such as UAuGe (Au-Au = 3.27 Å), which crystallizes in the YPtAs structure, and in ScAuSi (Au-Au = 2.94 Å) which has its own hexagonal structure type with Au-Au bonds (see Figure1).[29,30] Additionally, the Au-Au contacts in EuAuGe (Au-Au = 3.16 Å) are suggested to arise from realtivistic interactions.[31] Weak, secondary Au-Au and Sn-Sn interactions are also important driving factors for the formation of a $KHg_2$-type superstructure for YbAuSn (Au-Au = 3.32 Å).[32] Because the Au-Au interlayer contacts in $Ln$AuSb are even shorter than these (Au-Au = 2.98-3.12 Å), it is likely that the structure of the $Ln$AuSb compounds described here is driven by these gold-gold interactions.

**Electronic Structure Calculations**

To gain further insight into the bonding interactions in LaAuSb, we performed several types of electronic structure calculations. Figure 5 shows the density-of-states (DOS) curves with Au 6*s*, 5*d*, and 6*p* states highlighted, as well as the crystal orbital Hamilton population (COHP) data calculated (LMTO) for LaAuSb (see Table S4 for -ICOHP parameters). A prominent feature in the DOS is the strong suppression of states (pseudogap) at the Fermi level together with empty La-Au, La-Sb and Au-Au bonding states, while Au-Sb states are antibonding just below $E_F$. Within +/- 1eV of the indicated $E_F$, the overall COHP is nonbonding; the location of the $E_F$ is favored by the low DOS. This indicates that 19 valence electrons optimizes the bonding of the whole compound. Most of the DOS curve between -7 to -4 eV below the Fermi level belongs to the Au-5*d* and Au-6*s* electrons. The Au states (*s*, *p* and *d*) from -3.5 eV to 0 eV integrate to ~0.95 electrons per gold; this corresponds to the gold-gold interlayer bond. Above -3 eV, most contribution to the DOS comes from Sb-5*p* electrons and La 6*s* and 5*d* electrons (the *f*-states are treated as highly localized). This part of the DOS contains hybridized Au-Sb and La-Sb interactions according to the corresponding COHP curves, as one would expect for the covalently-bonded, hexagonal Au-Sb net.

Extended Hückel theory was then applied using Slater-type zeta functions. Figure 6 illustrates the highest occupied molecular orbitals (HOMOs) and lowest unoccupied molecular orbitals (LUMOs) for LaAuSb, which provide an interpretation for how the Au-Au dimers bond in LaAuSb. Those orbitals have contributions from Au 6*s*, 5*d*, and 6*p* orbitals, consistent with the DOS and COHP calculations discussed above. The electrons in the Au bond are more "molecular" than band-like in nature, in the sense that they reside primarily between the Au atoms. This dimerization provides electronic stability to LaAuSb through its facilitation of a nearly complete electronic band gap between filled and empty states. The valence electrons of LaAuSb are counted as 3 ($La^{3+}$) + 11 ($Au^0$) + 5 ($Sb^{3-}$) = 19-electrons, which would be unstable in a LiGaGe-type compound as antibonding states would be populated. However, that is not the case in the current structure type since one extra electron per Au is in the localized Au-Au dimers according to the integrated DOS; the remaining 18 electrons yield a nearly fully-gapped band structure similar to that seen in other 111 hexagonal phases. This family can therefore be viewed as $Ln^{3+}_2$(Au-Au)$^0$Sb$^{3-}_2$. Within the same (YPtAs) structure type several non-gold containing 19-electron systems are known to exist. These include $Ln^{3+}$ZnSn[33], $Ln^{3+}$ZnGe[34], and $Ln^{3+}$ZnPb[35]. The 19-electron germanide series shows no interlayer bonding interaction and the stannides show weak Zn-Zn bonding.[34] Our calculations illustrate that the electronic structures of YPtAs-type phases can best be interpreted using a combination of "molecular orbital" and band electron perspectives, and not just via the nearly free electron band model.

We now compare the ab-initio electronic band structure calculations for LaAuSn and LaAuSb generated using *WIEN2k*. Figure 7 shows both the band structures and the density of states (DOS) for LaAuSn and LaAuSb (see SI Figure S3 for similar calculations on LaAuPb). LaAuSn is a semimetal with a pseudogap in the electronic density of states at the Fermi level, straightforwardly consistent with its 18 valence electron count. This character can be simply understood by the fact that it is a charge-balanced compound; we therefore expect semiconducting or semimetallic (when the degree of covalency in the bonding is high and there is a nearly complete but not quite complete energy gap between the valence band and conduction band) behavior. If the Au atoms in $Ln$AuSb ($Ln$ = La-Nd, Sm) form dimers as described here, with the one extra electron accommodated in a localized Au-dimer orbital, then we expect a strong suppression of the density of states and a pseudo gap at the Fermi level since they will also be charge-balanced. From the calculations, we indeed find this to be the case, in agreement with the DOS calculations derived from the LMTO calculation. The appearance of the nearly fully gapped electronic structure of LaAuSb, where there is only one place in the Brillouin zone that is not gapped out, is therefore yet another indication for the presence of a true Au-Au bond between the layers. We note that there is bulk Dirac cone approximately 0.1 eV below the Fermi level in the Γ–A direction. This cone, along with the band crossing that creates a small DOS at $E_F$, is protected by the $C_3$ and $C_6$ symmetries along this line and cannot be gapped without a structural distortion to lower crystal symmetry or change in band overlap due to a significant change in lattice size.[36] Materials with bulk Dirac cones, or "Dirac Semimetals", have been of recent interest due to their exotic physical properties such as extremely high carrier mobility.[37]

**$Ln$AuZ Phase Comparisons.**

To help place the LnAuSb phases reported here in context with all other LnAuZ phases, we compiled the *Ln*AuX compounds known to exist as of the time of this publication. The results are summarized in Figure 8,[25, 12, 38, 39, 40, 41, 42, 43, 44, 45, 46, 47, 48, 49] where the Y-axis is the $Ln^{3+}$ ionic radius and the X-axis is the sum of the AuZ metallic radii, as has been done for other 111 compounds.[50] To remind the reader, LiGaGe and YPtAs are both buckled hexagonal structures, ZrBeSi is a 2-layer unbuckled hexagonal structure, MgAgAs is the prototypical cubic Half-Heusler structure, and the $KHg_2$-type compounds are orthorhombically-distorted superstructures of stacked honeycombs[31, 32, 51].

Several observations can be made. First, the $Ln^{3+}$AuSb systems are the only *Ln*AuZ phases known to crystallize in the YPtAs structure type, and there appear to be no other 19-electron LnAuZ systems. The *Ln*AuGe, *Ln*AuSn, and *Ln*AuPb compounds, of which there are many, are all 18-electron systems for $Ln^{3+}$. Of the remaining *Ln*AuAs, *Ln*AuSb, and *Ln*AuBi phases the only phases known to exist besides the $Ln^{3+}$AuSb compounds reported here are those based on the divalent rare earths (Yb and Eu), which have an 18-electron count (dashed line in Figure 8). These 18-electron phases crystallize in the LiGaGe type rather than the YPtAs type, as would be expected, since no interlayer dimer is required to hold an extra electron. Therefore, to the best of our knowledge, the YPtAs-type *Ln*AuSb reported here are a unique, 19-electron group of *Ln*AuZ phases.

We also observe that there is a nearly linear phase boundary between the hexagonal phases (YPtAs, LiGaGe, ZrBeSi) and the cubic Half-Heusler phase. This phase boundary is consistent with that observed by Xie, et al[50]; the current *Ln*AuSb system develops further the 111 structural stability diagram specific to the rare earth cation radii. To clarify the phase boundary in greater detail, we attempted to synthesize several other 111 phases, including $Ln$AuSb ($Ln$ = Gd, Tb, Dy, Tm, Sc, and Y) and $Ln$AuBi ($Ln$ = La, Tm), but none were found to exist within our reaction conditions. Thus the boundary between the hexagonal and Half Heusler structure is just below $Tb^{3+}$ for $Ln$AuSb; while this does not explain why we could not successfully synthesize "GdAuSb", it does explain why smaller-sized rare earth variants were not found to exist in a hexagonal structure phase. There appears no straightforward way to incorporate Au dimers into a variant of the Half Heusler structure to obtain a 19-electron system, explaining why $Ln^{3+}$AuSb and $Ln^{3+}$AuBi phases are not found in the Half-Heusler structure type. The phase boundary in the figure suggests that hexagonal $Ln^{3+}$AuBi compounds may not exist.

We infer that it may be possible to synthesize yet-undiscovered arsenide equivalents of the current phases, i.e. $Ln^{3+}$AuAs phases, which would have 19 valence electrons. We believe these would have the YPtAs-type structure. Given all the interesting properties arising arsenides recently, these may be of interest for future work. Should the reader be interested in synthesizing these phases, caution is advised as As is toxic and requires care and precautions for handling.

Finally, we observe the general trend that for compounds that exist as both the hexagonal and cubic variants (for example HoAuSn[39] and YbAuBi[43]), the hexagonal phase is the high-temperature phase and the cubic phase is the low-temperature phase. Though the $Ln$AuSb compounds reported here were not found to be polymorphic, this observation supports the notion that these are high-temperature phases that require rapid quenching and may have competing polymorphs at low temperatures. It also suggests that if $Ln^{3+}$AuAs were to be made they may be high-temperature phases.

**Conclusions**

We find that the new $Ln$AuSb ($Ln$ = La, Ce, Nd) compounds crystallize in the YPtAs structure type via single crystal diffraction and high-resolution powder diffraction data. We also find evidence for PrAuSb and SmAuSb crystallizing in the same structure type, and report preliminary lattice parameters for those compounds. The structure of LaAuSn was reinvestigated and found to be the ordered LiGaGe-type. The $Ln$AuSb ($Ln$ = La – Nd, Sm) phases are a more buckled version of the archetypical YPtAs structure, allowing for the formation of interlayer Au-Au dimers. These dimers localize the "19$^{th}$" electron and are important in maintaining a stable structure. Importantly, this dimerization causes the formation of a YPtAs structure type, which allows for band crossing at the Fermi level. Ab-initio band structure calculations indicate that these materials are semimetals with an electronic band gap over nearly the full Brillouin zone, with a bulk Dirac cone along Γ-A. We propose that these materials may therefore be of interest for further study, and predict the stability of currently unobserved $Ln^{3+}$AuAs phases. We encourage readers to consider new ways in which the tendency for gold-gold bonding may stabilize previously unrealized compounds with new and unique properties.

**Acknowledgements**

This research was supported by the Air Force MURI on thermoelectric materials, Grant No. FA9550-10-1-0553, and by SPAWAR grant NN66001-11-1-4110. The authors would like to thank Dr. Jing Gu for helpful contributions.

**Supporting Information Available**

Powder diffraction refinements for CeAuSb and NdAuSb (Figures S1 and S2) and selected bond lengths for LaAuSb, CeAuSb, and NdAuSb (Table S3); single crystal refinement parameters for LaAuSn (Table S1) and atomic coordinates (Table S2); Band structure and DOS of LaAuPb (Figure S3); -ICOHP parameters for LaAuSb (Table S4); cif outputs. This information is available free of charge via the internet at http://pubs.acs.org.

**Corresponding Authors**

*eseibel@princeton.edu

#rcava@princeton.edu

**Figure Captions**

**Figure 1**. The crystal structures of the archetype hexagonal 111- compounds LiGaGe, ZrBeSi, ScAuSi, and YPtAs. The top panel illustrates the stacking along $c$; the bottom panel demonstrates the honeycomb $Y_3Z_3$ nets in the $a$-$b$ plane. LiGaGe and ZrBeSi are 2-layer honeycomb structures that stack with alternating *YZYZ* atoms along $c$, but LiGaGe has buckled $Y_3Z_3$ honeycomb layers whereas ZrBeSi is flat. ScAuSi is also a buckled, 2-layer structure, but with *YYYY* stacking to allow for interlayer *Y-Y* contacts. YPtAs is a slightly buckled 4-layer structure with *YYZZ*-type stacking along c, such that extensive buckling of the $Y_3Z_3$ layers could allow for *Y-Y* interlayer bonds.

**Figure 2.** Rietveld refinement of LaAuSb. Observed synchrotron powder X-ray diffraction data is shown in red, calculated in black, and the difference ($Y_{obs}$-$Y_{calc}$) in blue. The insets show the peak shapes and fit to the data from 15-20° $2\theta$. Green tick marks are Bragg reflections for LaAuSb (top), $La_3Au_2Sb_3$ (middle), and $La_3Au_3Sb_4$ (bottom).

**Figure 3.** The structure of LaAuSb. CeAuSb and NdAuSb are isostructural; we assume the same for PrAuSb and SmAuSb based on their crystallographic cell parameters. The $Au_3Sb_3$ hexagonal layers are easily visible in *a-b* projection, shown in the top left. The top right shows the projection of the *b-c* plane. Au-Sb and Au-Au bond lengths are indicated on the figure. The bottom portion shows the coordination polyhedra for Au, Sb, and the 2 La sites. If the Au-Au dimer is treated as a unit, the dumbbell adopts 9-fold coordination formed by a Sb trigonal prism and trigonal planar La bonds. Sb has trigonal prismatic coordination with La through which there are three bent Au bonds. Both La display dodecahedral coordination, though with different Au and Sb at the vertices.

**Figure 4**. Structure comparisons of YPtAs (left) and LaAuSb (right). Although LaAuSb adopts the YPtAs structure type, the Au and Sb atoms are significantly more buckled than the Pt and As atoms in the archetypical structure, which is a signature of the Au-Au interlayer bonding. The black arrows in the figure are meant to indicate the direction of motion of the As site to allow for interlayer Au-Au bond formation.

**Figure 5.** The DOS and COHP for LaAuSb calculated using LMTO. The gold 6$s$, 5$d$, and 6$p$ states that contribute to the total DOS are highlighted on the left. La1-Sb, La2-Au, Au-Sb, and Au-Au interactions are highlighted in the COHP, shown on right. The gold 6$s$ and 5$d$ states are highly localized between -4 and -6.5 eV in the DOS plot, which creates bonding interactions (-6.5 to -5 eV) and antibonding interactions (-5 to -3 eV) seen in the COHP. The overall bonding interactions from approximately -4 eV to -1.5 eV are made of hybridized Au, Sb, and La states. The compound is nonbonding in the COHP from +/-1 eV around the indicated Fermi level, which sits in a deep depression of the density of states, a "pseudogap".

**Figure 6**. The HOMOs and LUMOs of LaAuSb calculated using extended Hückel theory with relativistic effects included. The unit cell of LaAuSb is shown for comparison with Au-Au bonds; the red outline indicates the part of the crystal structure shown in the MO calculations. The sign of the wavefunction is indicated by red and blue color. The strong orbital overlap between interlayer Au atoms creates a bond in which two electrons (one from each Au) are localized.

**Figure 7** Ab-initio electronic band structures (spin orbit coupling included) and density of states (DOS) of LaAuSn (a) and LaAuSb (b). The compounds have semimetallic electronic structures, with nearly a full band gap between occupied and empty states. LaAuSb (b) has a bulk Dirac cone approximately 0.1 eV below the Fermi level in the Γ–A direction that is symmetry-protected; this protection gives rise to the small valence band-conduction band overlap near the gamma point in the Brillouin Zone.

**Figure 8**. The structural stability phase diagram of $Ln$AuZ phases (Z = Ge, As, Sn, Sb, Pb, Bi). The plot is an analogy to Ref. 49. The LiGaGe structure type is marked with blue squares, the YPtAs type with light blue stars, ZrBeSi type with teal triangles, MgAgAs (Half Heusler) with red circles, and KHg$_2$-type with green squares. Red shading indicates a region of cubic symmetry (i.e. Half Heusler) whereas blue

shading indicates a region of hexagonal symmetry. There are several polymorphic phases that fall within a purple region. The dashed line serves as a guide to the eye for $Yb^{2+}$ and $Eu^{2+}$ phases, which often crystallize in a structure type that differs from the rest of the $Ln^{3+}$AuZ family. Each LnAuZ column is indicated as "18e-" or "19e-" based on the counting scheme adopted in the text for $Ln^{3+}$. "HT" and "LT" stand for high-temperature and low-temperature phases, respectively. There is a clear boundary between the hexagonal and cubic 111 phases, as well as an absence of LnAuAs phases.

**Table 1. Crystal Data and Structure Refinements for LnAuSb (Ln = La-Nd, Sm)**

| Formula | LaAuSb[#] | CeAuSb[#] | PrAuSb[*] | NdAuSb[#] | SmAuSb[*] |
|---|---|---|---|---|---|
| Formula Weight (g/mol) | 457.63 | 458.84 | 459.63 | 462.97 | 469.09 |
| Space Group | $P6_3/mmc$ | $P6_3/mmc$ | $P6_3/mmc$ | $P6_3/mmc$ | $P6_3/mmc$ |
| Z | 2 | 2 | 2 | 2 | 2 |
| Unit Cell (Å) | | | | | |
| $a$ | 4.63838(6) | 4.6140(1) | 4.593(2) | 4.5800(1) | 4.551(2) |
| $c$ | 16.8315(4) | 16.6348(6) | 16.532(1) | 16.4775(5) | 16.398(1) |
| Volume | 313.60(1) | 306.66(1) | 302.13(3) | 299.34(1) | 294.21(3) |
| $\chi^2$ | 3.00 | 4.91 | 3.05 | 3.92 | 2.67 |
| $R_{wp}$ | 14.4 | 13.5 | 28.2 | 13.8 | 26.4 |
| $R_p$ | 13.9 | 12.4 | 26.6 | 12.7 | 23.5 |
| Impurity phases | $La_3Au_2Sb_3$ $La_3Au_3Sb_4$ | $Ce_3Au_2Sb_3$ $Ce_{14}Au_{51}$ | $Pr_3Au_2Sb_3$ $Pr_{14}Au_{51}$ | $Nd_3Au_2Sb_3$ $Nd_{14}Au_{51}$ | $Sm_3Au_3Sb_4$ $Sm_{14}Au_{51}$ |
| # Rietveld refinement from synchrotron data  * Profile fit from lab PXRD ||||||

**Table 2.** Atomic coordinates and thermal parameters for LnAuSb (Ln = La, Ce, Nd) phases

| Phase | Atom | Wyckoff | $x$ | $y$ | $z$ | $B_{iso}$ | Occ. |
|---|---|---|---|---|---|---|---|
| LaAuSb | La1 | *2a* | 0 | 0 | 0 | 0.86(6) | 1 |
| | La2 | *2b* | 0 | 0 | 1/4 | 0.86(6) | 1 |
| | Au1 | *4f* | 2/3 | 1/3 | 0.1572(1) | 1.34(4) | 1 |
| | Sb1 | *4f* | 1/3 | 2/3 | 0.1127(2) | 0.78(7) | 1 |
| CeAuSb | Ce1 | *2a* | 0 | 0 | 0 | 0.9(1) | 1 |
| | Ce2 | *2b* | 0 | 0 | 1/4 | 0.9(1) | 1 |
| | Au1 | *4f* | 2/3 | 1/3 | 0.1583(2) | 1.24(7) | 1 |
| | Sb1 | *4f* | 1/3 | 2/3 | 0.1122(4) | 0.8(1) | 1 |
| NdAuSb | Nd1 | *2a* | 0 | 0 | 0 | 0.8(1) | 1 |
| | Nd2 | *2b* | 0 | 0 | 1/4 | 0.8(1) | 1 |
| | Au1 | *4f* | 2/3 | 1/3 | 0.1596(2) | 1.09(7) | 1 |
| | Sb1 | *4f* | 1 | 1/3 | 2/3 | 0.8(1) | 1 |

**Figure 1**

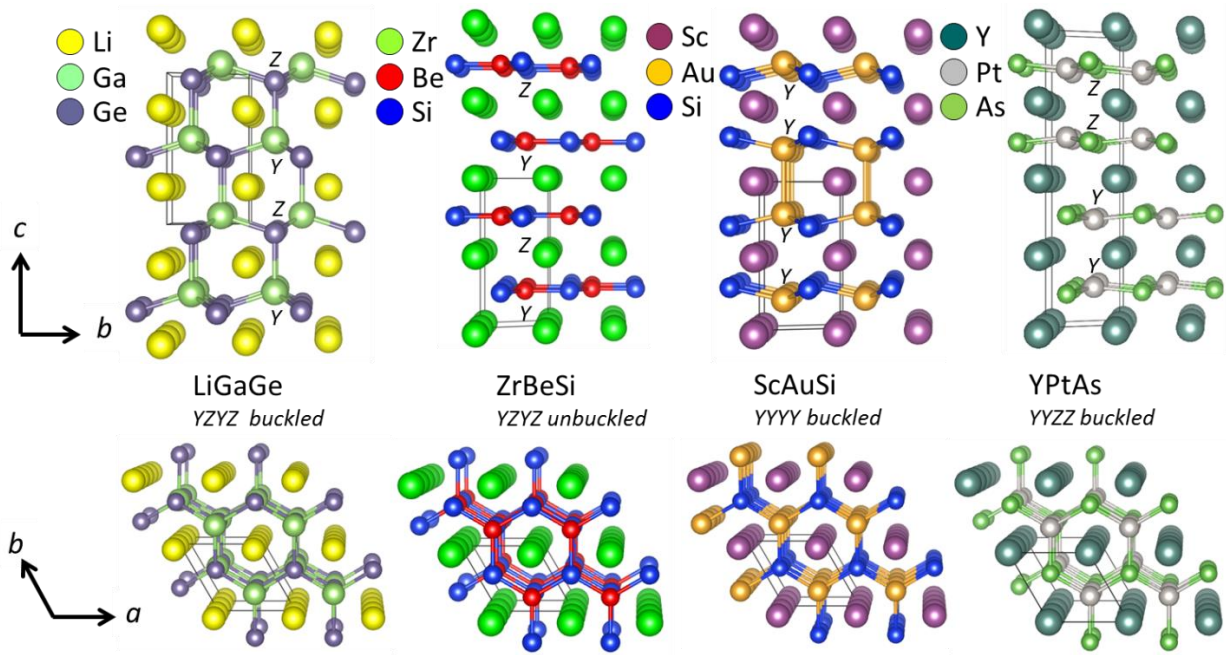

LiGaGe
*YZYZ buckled*

ZrBeSi
*YZYZ unbuckled*

ScAuSi
*YYYY buckled*

YPtAs
*YYZZ buckled*

**Figure 2**

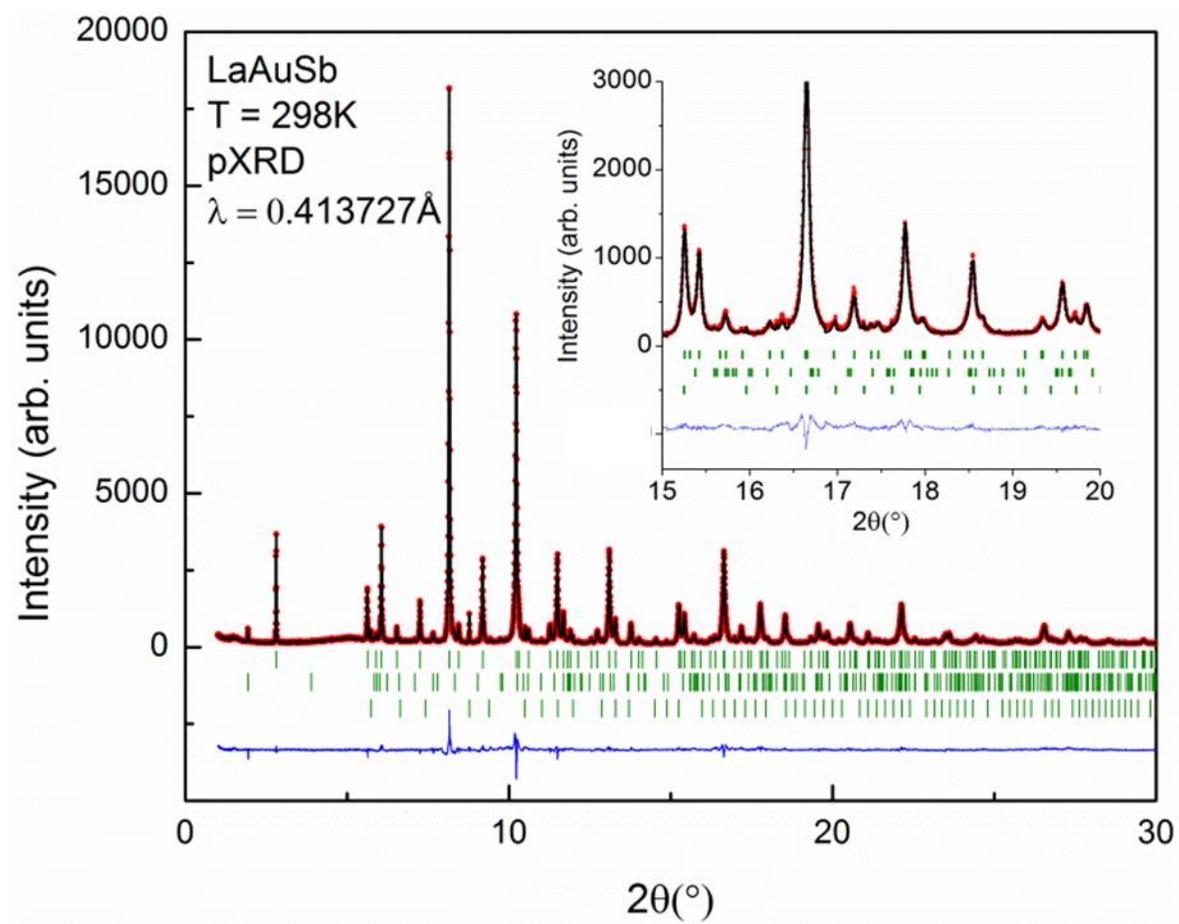

**Figure 3**

LaAuSb
- La
- Au
- Sb

3.12Å
2.78Å
2.78Å

La1
La2
La1

Au1  Sb1  La1  La2

**Figure 4**

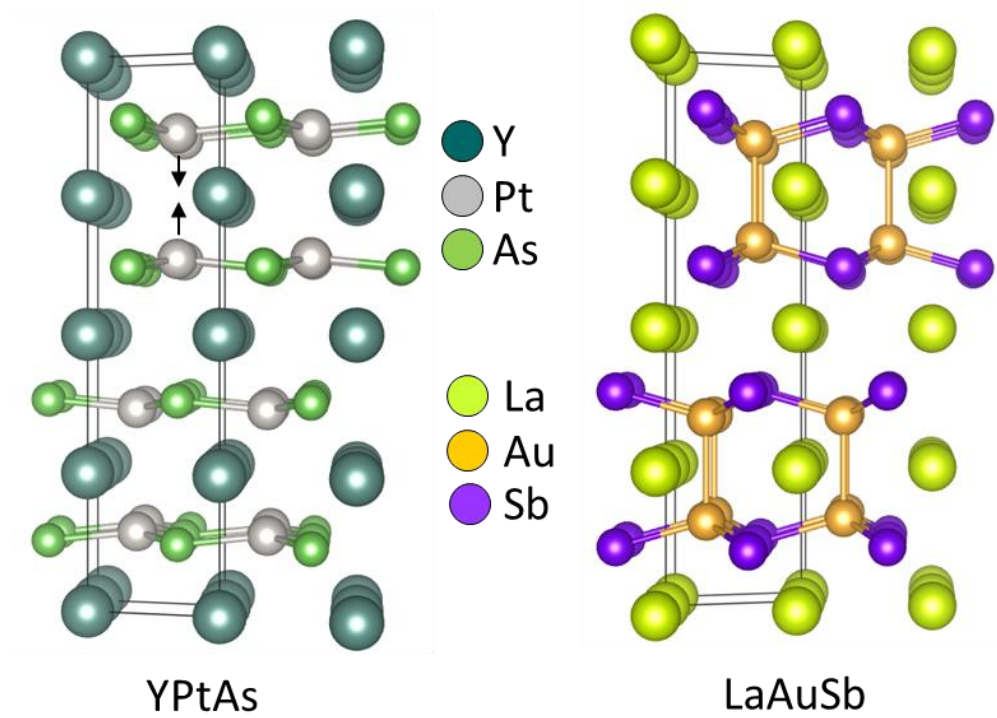

YPtAs  LaAuSb

**Figure 5**

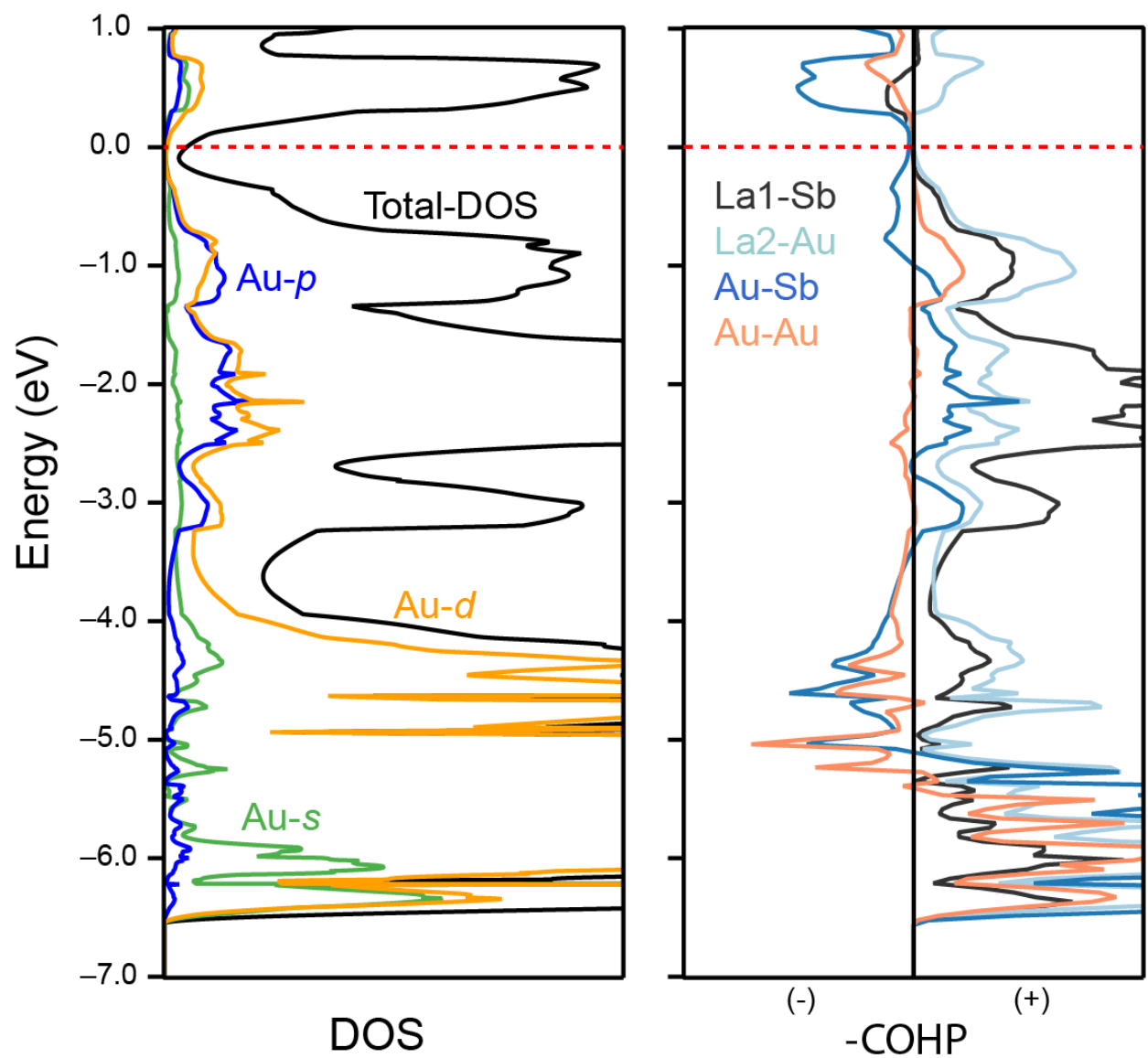

**Figure 6**

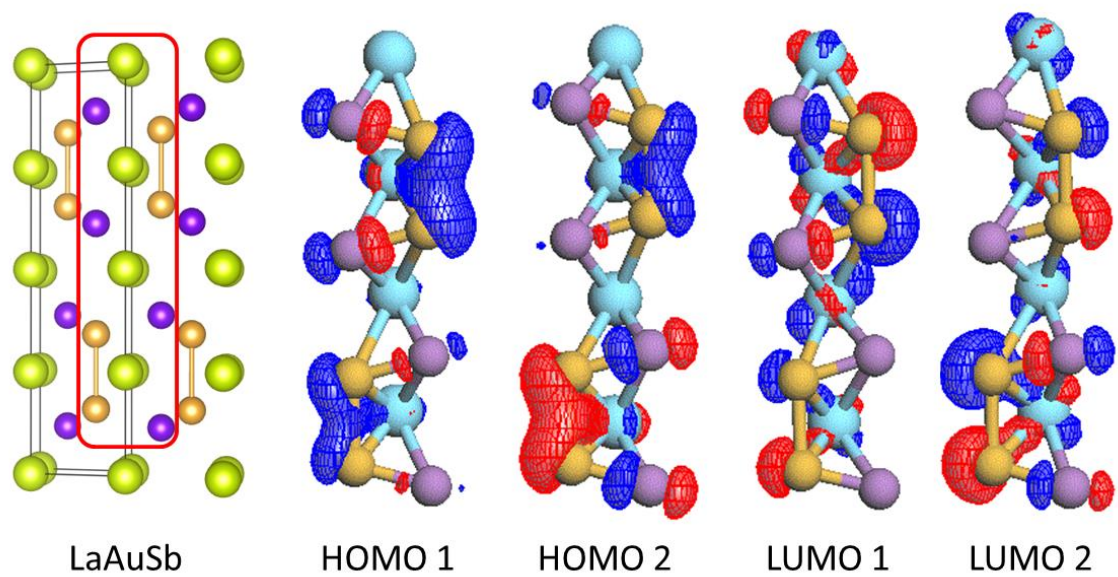

LaAuSb     HOMO 1     HOMO 2     LUMO 1     LUMO 2

**Figure 7**

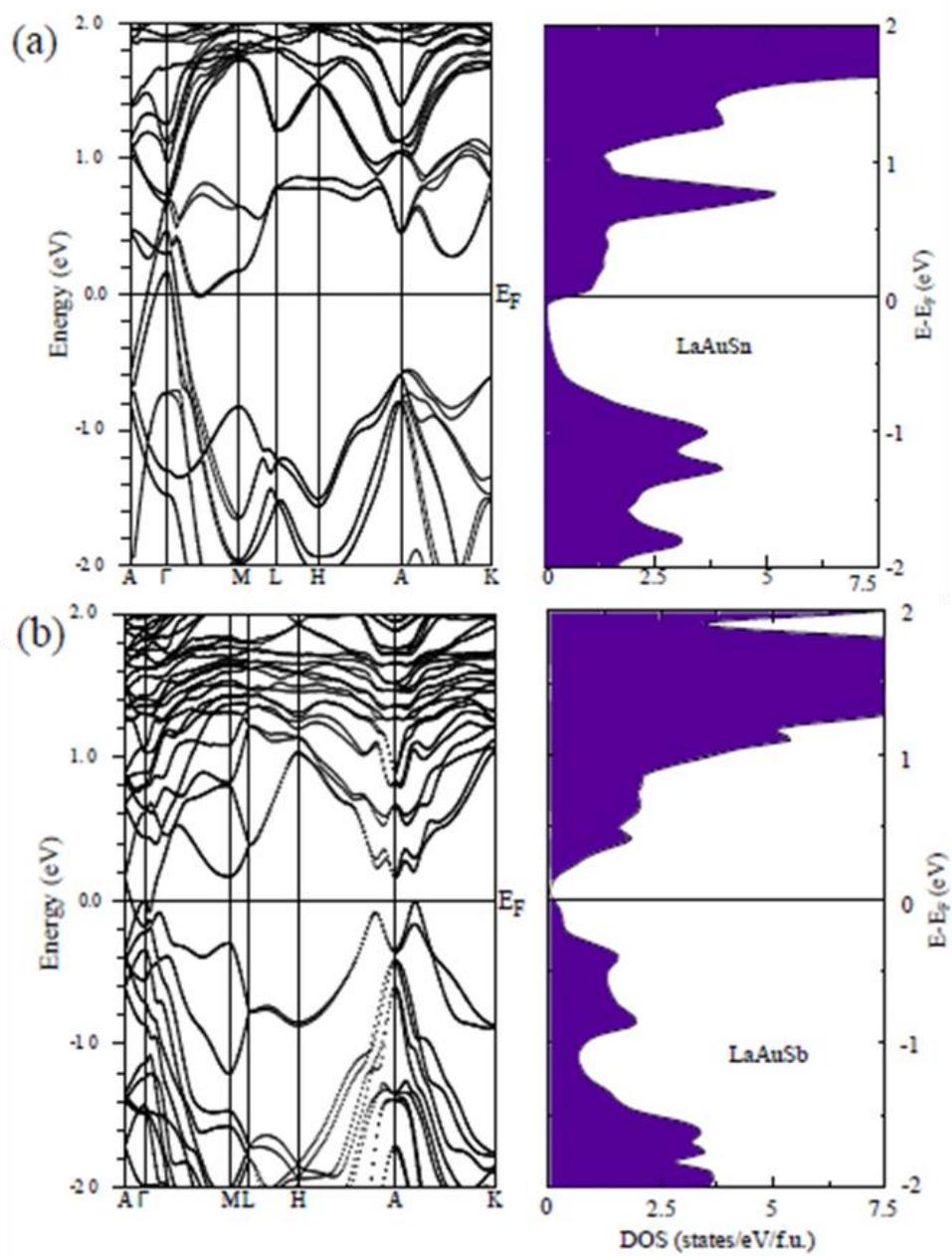

**Figure 8**

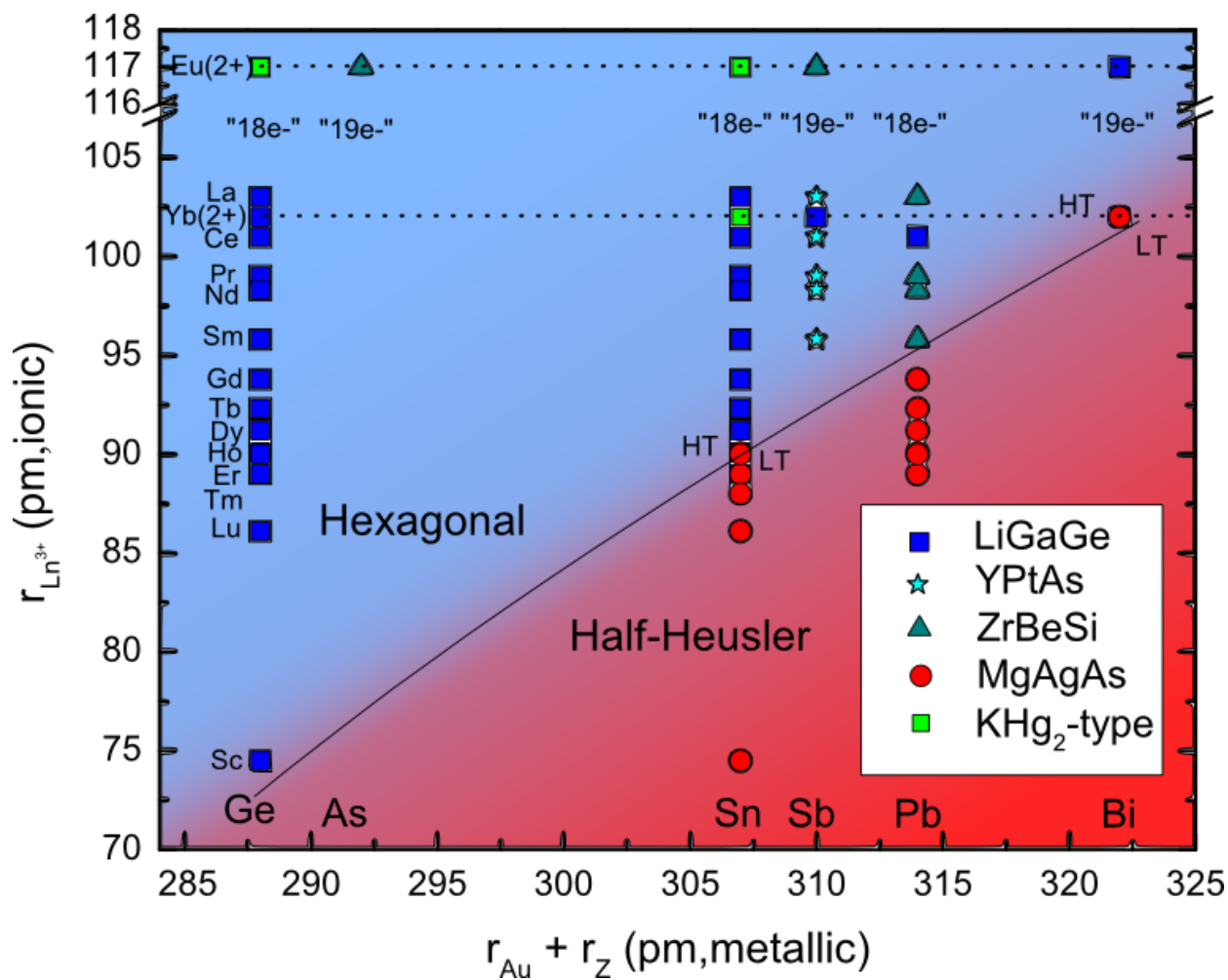

*Supporting Information*

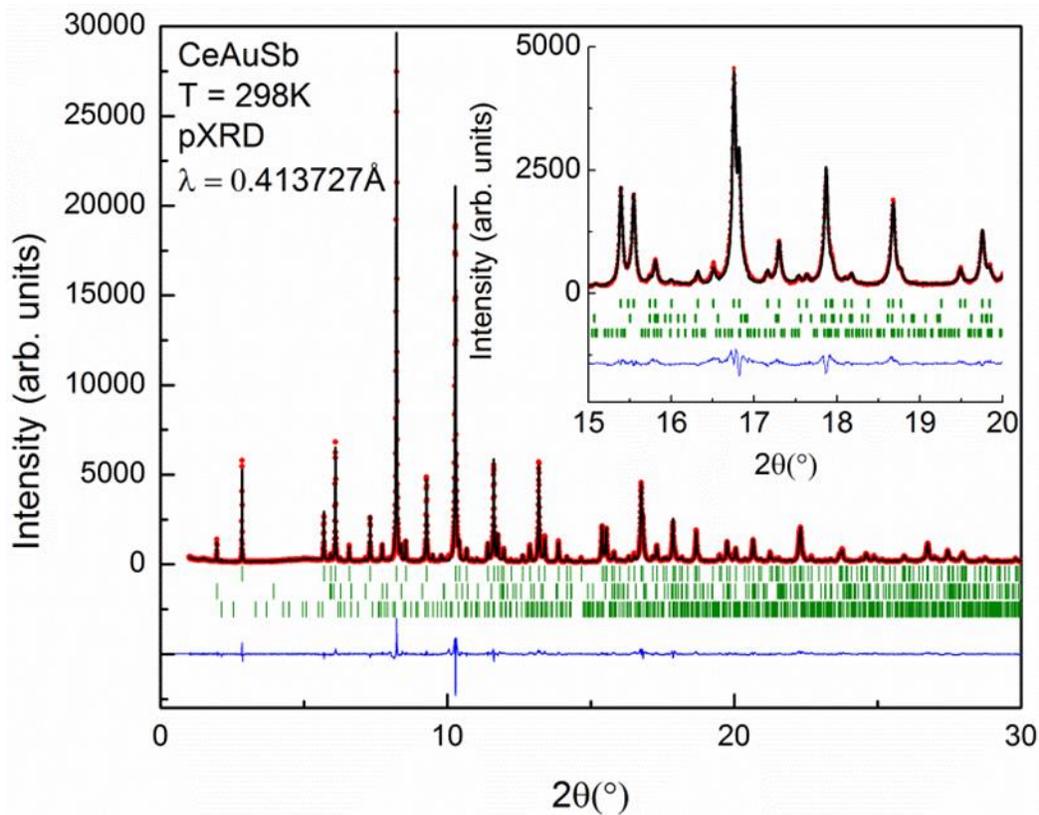

**Figure S1.** Rietveld refinement of CeAuSb. Observed synchrotron powder X-ray diffraction data is shown in red, calculated in black, and the difference ($Y_{obs}$-$Y_{calc}$) in blue. The insets show the peak shapes and fit to the data from 15-20° $2\theta$. B) Green tick marks are Bragg reflections for CeAuSb (top), $Ce_3Au_2Sb_3$ (middle), and $Ce_{14}Au_{51}$ (bottom).

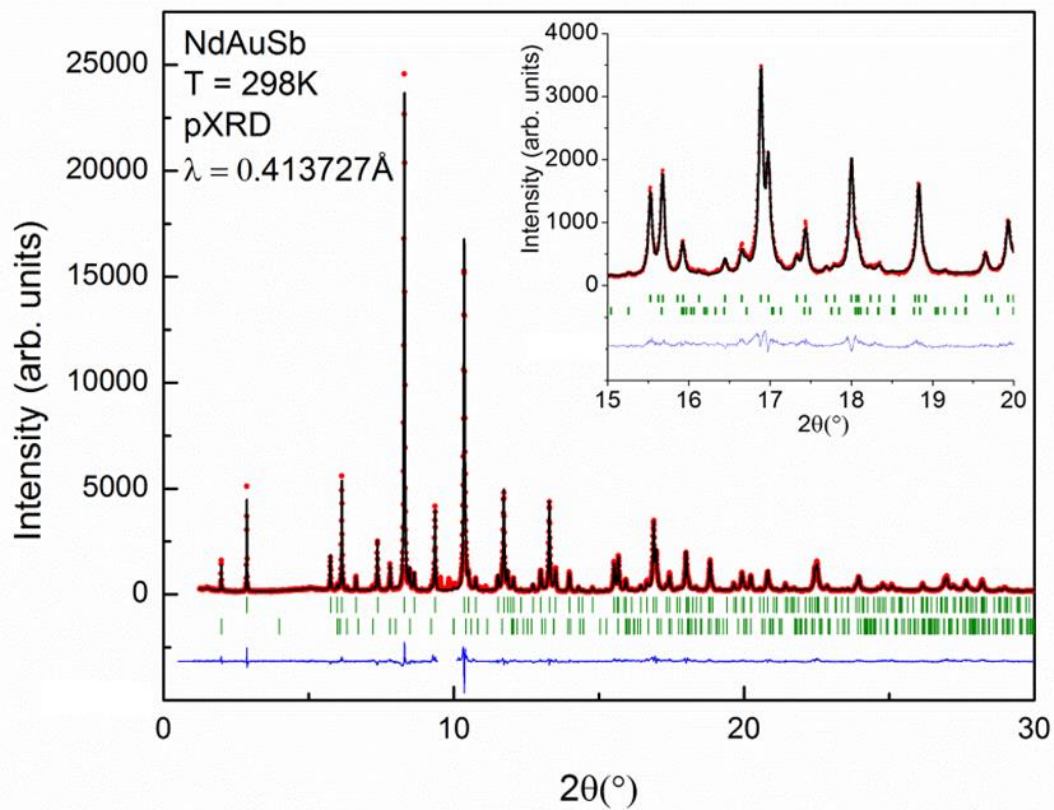

**Figure S2**. Rietveld refinement NdAuSb. Observed synchrotron powder X-ray diffraction data is shown in red, calculated in black, and the difference ($Y_{obs}-Y_{calc}$) in blue. The insets show the peak shapes and fit to the data from 15-20° $2\theta$. Green tick marks are Bragg reflections for NdAuSb (top) and $Nd_3Au_2Sb_3$ (bottom). Note that impurity $Nd_{14}Au_{51}$ peaks were omitted from the refinement (~9-10° $2\theta$) due to excessive peak overlap.

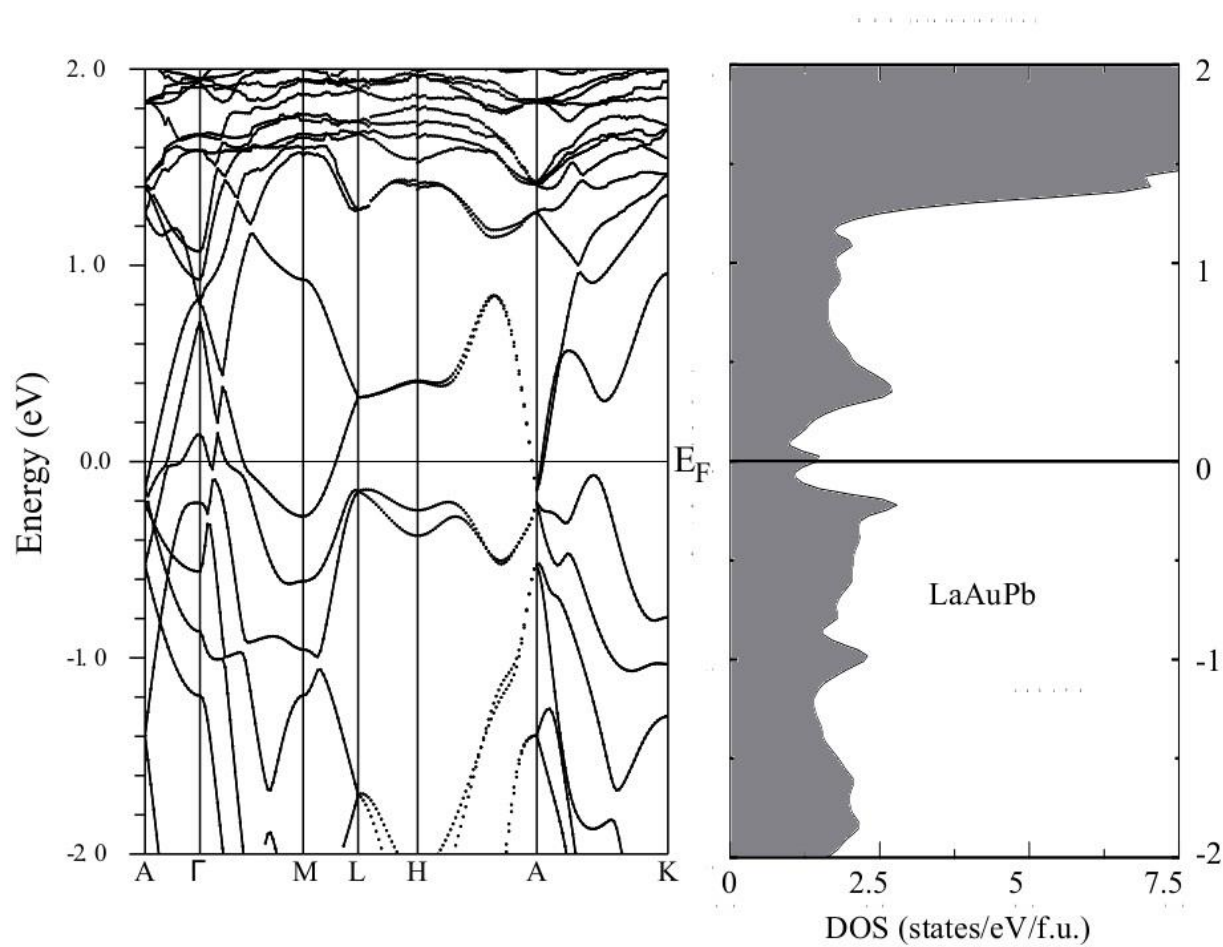

**Figure S3.** Band structure and DOS of LaAuPb. For the electronic structure calculations on LaAuPb, we noticed that the reported[28] ZrBeSi structure type (the non-buckled version of the LiGaGe structure type, Figure 1) may not be correct. In this case the calculated DOS (not shown) is significantly higher than that in the other materials. This high DOS hints that the compound may actually crystallize in the buckled LiGaGe structure, where the DOS would be lower, and not in the reported, unbuckled structure type. In addition, there is a van Hove singularity on the Γ-A line, which also suggests that the ZrBeSi structure type should be unstable for LaAuPb. Finally, unlike LaAuPb, CeAuPb has been reported to crystallize in the buckled LiGaGe structure[7]; this also suggests that the LaAuPb structure has been described incorrectly and should in fact have buckled honeycomb layers.

**Table S1.** Single crystal crystallographic data for LaAuSn phases at 296(2) K.

| Loading composition | LaAuSn |
|---|---|
| Refined Formula | La$_2$Au$_2$Sn$_2$ |
| F.W. (g/mol); | 909.13 |
| Space group; Z | P6$_3$mc(No.186); 1 |
| a (Å) | 4.7380(8) |
| c (Å) | 7.863(1) |
| V (Å$^3$) | 152.87(6) |
| Absorption Correction | Multi-Scan |
| Extinction Coefficient | 0.0048(6) |
| μ(mm$^{-1}$) | 69.354 |
| θ range (deg) | 4.968-28.022 |
| hkl ranges | −6≤ h,k ≤ 6<br>−10≤ l ≤ 10 |
| No. reflections; $R_{int}$ | 1347; 0.0147 |
| No. independent reflections | 175 |
| No. parameters | 12 |
| $R_1$; $wR_2$ (all I) | 0.0117; 0.0180 |
| Goodness of fit | 1.289 |
| Diffraction peak and hole (e$^-$/Å$^3$) | 0.385; −0.713 |

**Table S2.** Atomic coordinates and equivalent isotropic displacement parameters of LaAuSn. U$_{eq}$ is defined as one-third of the trace of the orthogonalized U$_{ij}$ tensor (Å$^2$).

| Atom | Wyckoff. | Occupancy. | x | y | z | U$_{eq}$ |
|---|---|---|---|---|---|---|
| La | 2a | 1 | 0 | 0 | 0.8174(1) | 0.0074(2) |
| Au | 2b | 1 | 1/3 | 2/3 | 0.4801(1) | 0.0132(2) |
| Sn | 2b | 1 | 1/3 | 2/3 | 0.0889(1) | 0.0072(2) |

**Table S3.** Selected bond lengths for LaAuSb, CeAuSb, and NdAuSb

|  | LaAuSb | CeAuSb | NdAuSb |
|---|---|---|---|
| Au1 – Au1 (x1) | 3.116(1) | 3.0455(9) | 2.976(1) |
| Au1 – Sb1 (x3) | 2.7814(4) | 2.7715(4) | 2.7664(5) |
| Au1 – Ln2 (x3) | 3.0984(4) | 3.0684(3) | 3.0343(4) |
|  |  |  |  |
| Sb1 – Au1 (x3) | 2.7812(4) | 2.7717(4) | 2.7661(5) |
| Sb1 – Ln1 (x3) | 3.2827(8) | 3.2551(7) | 3.2093(8) |
| Sb1 – Ln2 (x3) | 3.5361(9) | 3.5116(8) | 3.505(1) |

**Table S4.** –ICOHP parameters for LaAuSb

| Atom I – Atom II | Distance (Å) × # | -ICOHP | % (-COHP) |
|---|---|---|---|
| La1-Sb | 3.282 × 12 | 0.9112 | 28.15 |
| La2-Au | 3.100 × 12 | 0.9246 | 28.57 |
| Au-Sb | 2.781 × 8 | 1.8713 | 38.54 |
| Au-Au | 3.122 × 2 | 0.9197 | 4.74 |

The parameters for Au are 6$s$: $\zeta$ = 1.890, $H_{ii}$ = –8.23 eV; 6$p$: $\zeta$ = 1.835, $H_{ii}$ = –4.89 eV, and 5$d$: $\zeta$ = 3.560, $H_{ii}$ = –12.200 eV. The Au parameters were modified to provide the best fit to the results of first-principles calculations with relativistic effect.